\pdfoutput=1
\documentclass[structabstract]{aa}  
\usepackage{graphicx}
\usepackage{txfonts}
\usepackage{natbib}

\newcommand{\Ms}{\ensuremath{M_{\odot}}}
\newcommand{\eg}{{\it e.g.}}
\newcommand{\cf}{{\it c.f. }}
\newcommand{\ie}{{\it i.e.}}

\newcommand{\xtr}{\ensuremath{\xi_{\rm true}}}
\newcommand{\fej}{\ensuremath{f_{\rm ej}}}
\begin{document}
\title{On the true shape of the upper end of the stellar initial mass function}

\subtitle{The case of R136}

\author{
        Sambaran Banerjee\inst{1}
        \and
        Pavel Kroupa\inst{1}
        }

\institute{
           $^1$Argelander-Institut f\"ur Astronomie,
           Auf dem H\"ugel 71, D-53121, Bonn, Germany\\
           \email{sambaran@astro.uni-bonn.de; pavel@astro.uni-bonn.de}
          }

\date{Received~~~~~~~~~~~~~~~; accepted~~~~~~~~~~~~~~~}

\abstract{The shape of the stellar initial mass function (IMF) of a star cluster near its upper mass limit
is a focal topic of investigation as it determines the high mass stellar content and hence the dynamics
of the cluster at its embedded phase as well as during its young gas-free phase. The massive stellar content
of a young cluster, however, can be substantially modified due to the dynamical ejections of the
massive stars so that the present-day high-mass stellar mass function (hereafter MF) can be different than that
with which the cluster is born.}
{In the present study, we provide a preliminary estimate of this evolution of the high-mass IMF
of a young cluster due to early ejections of massive stars, using the Large Magellanic Cloud massive, young
cluster R136 as an example.}
{To that end, we utilize the results of the state-of-the-art calculations by Banerjee, Kroupa \& Oh (2012)
comprising direct N-body computations of realistic, binary-rich, mass-segregated models of R136.
In particular, these calculations provide the ejection fraction of stars as a function of stellar mass.}
{We find that if the measured IMF of R136 is granted to be canonical, as observations
indicate, then the ``true'' high-mass IMF of R136 at its birth must be at least moderately top-heavy when
corrected for the dynamical escape of massive stars.}
{The top-heaviness of the true high-mass IMF over the observationally determined one is a general feature of massive,
young clusters where the dynamical ejection of massive stars is efficient. We discuss its implications
and possible improvements over our current estimate.}


\maketitle

%

\section{Introduction}\label{intro}

\begin{figure*}
\centering
\includegraphics[height=10cm,angle=0]{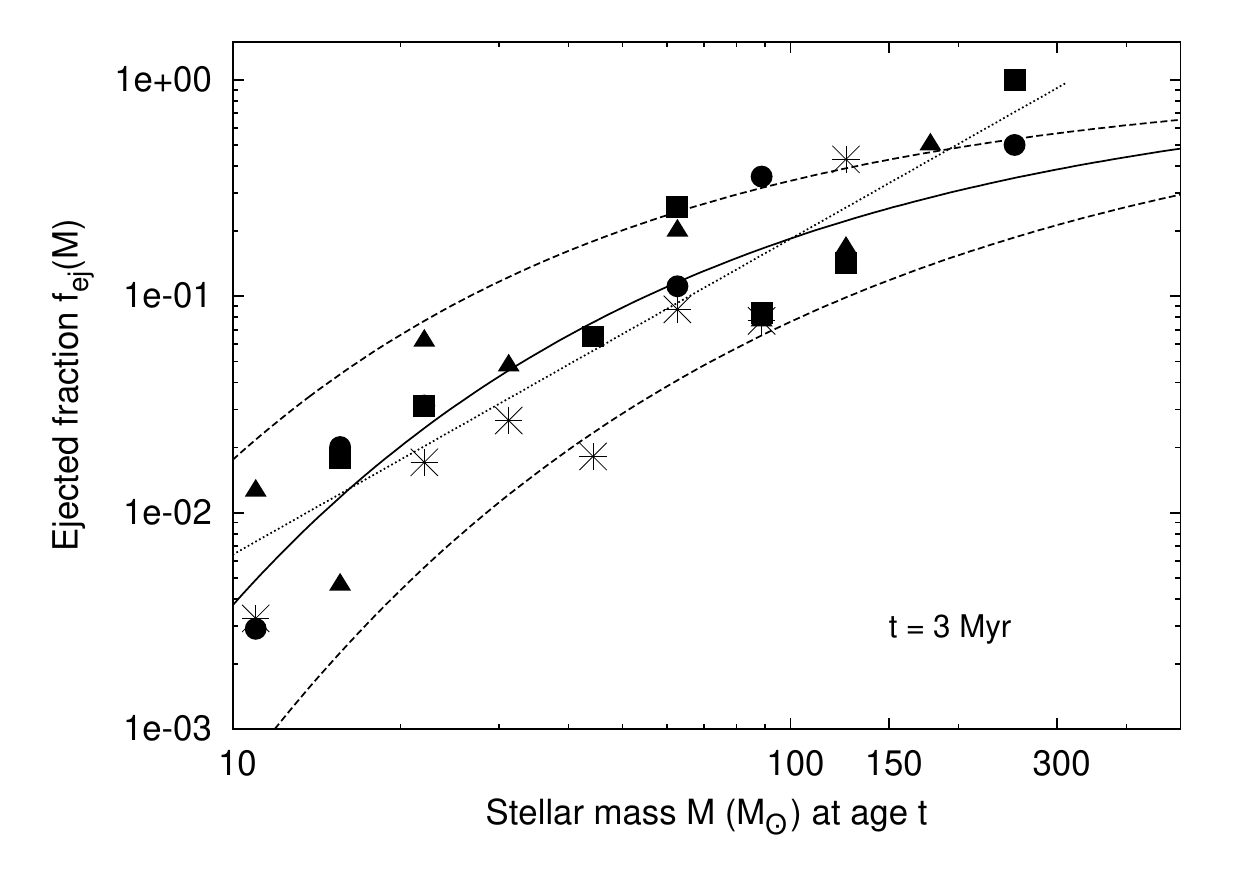}
\caption{The fraction $\fej(M)$ of runaway stars for $M>10\Ms$ at $t\approx 3$ Myr cluster age 
as obtained from the four computations of Paper I
(\cf Fig.~4 of Paper I). We use the same symbols as in Fig.~4 of Paper I to indicate that
they are obtained from direct N-body computations starting with different initial realizations of the R136 model
employed in Paper I. The central solid line represents the best-fit line to the data
that asymptotes to unity (see Sec.~\ref{ejfrac}). The upper and the lower dashed lines limit the error
in the fitting. The dotted line is the best-fit single power-law to the entire data which is truncated at
$f_{ej}(M)=1$ (see Sec.~\ref{ejfrac}).}
\label{fig:ejfrac1}
\end{figure*}

\begin{figure*}
\centering
\includegraphics[height=10cm,angle=0]{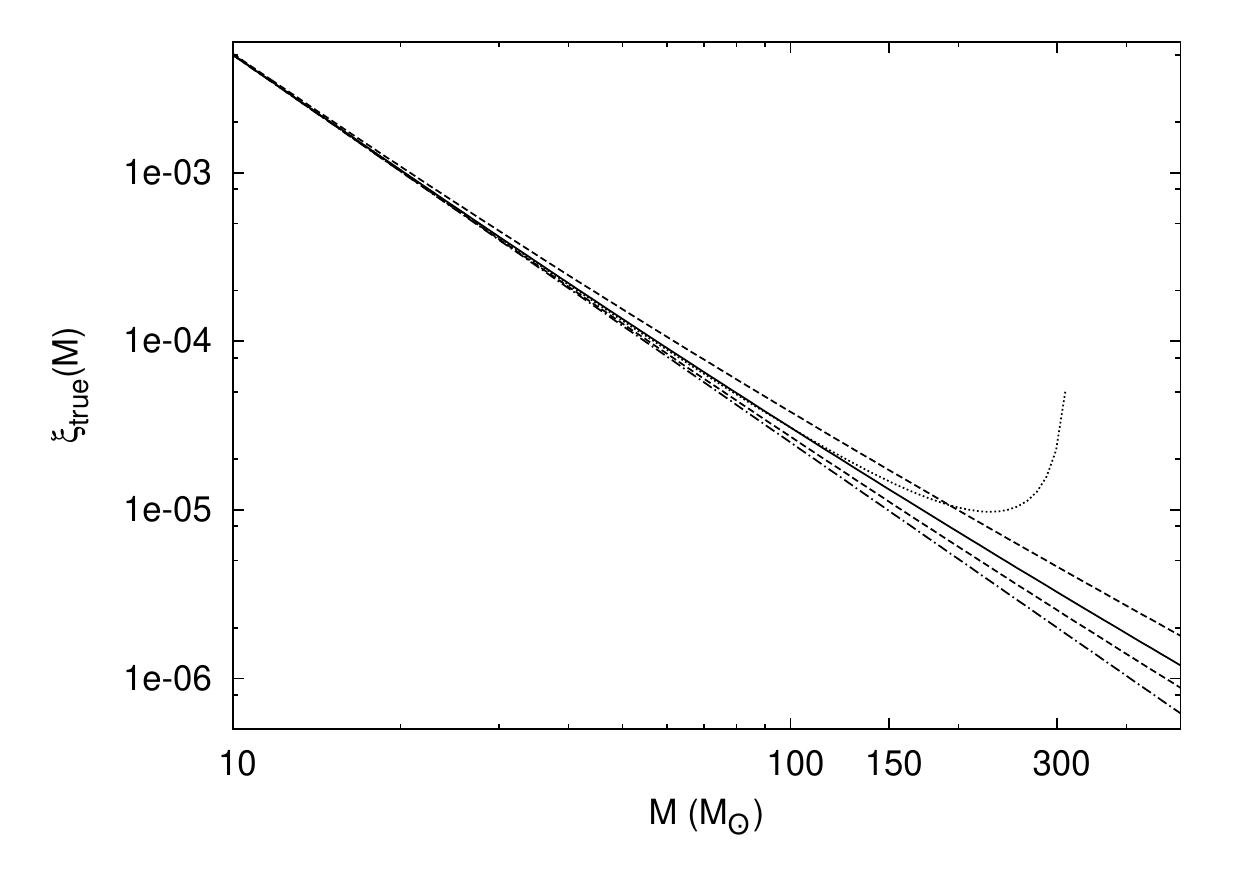}
\caption{The ``true'' IMF, $\xtr(M)$, as corrected for the ejected stars assuming the
``observed'' IMF, $\xi(M)$, to be the
Salpeter/canonical power law $\xi(M)\propto M^{-2.3}$ (represented by the dash-dotted line).
The true IMF is related
to the observed IMF by Eqn.~\ref{eq:phirho}.
The upper dashed, the lower dashed, the central solid and the dotted lines correspond to
those in Fig.~\ref{fig:ejfrac1} respectively.
It can be noted that the correction of
the observed IMF due to the dynamical ejection of massive stars can amount to
a moderately top-heavy true IMF (the dashed and the solid lines). However,
the data in Fig.~\ref{fig:ejfrac1} is also supportive of a substantially higher top-heaviness (the
dotted line).}
\label{fig:newMF}
\end{figure*}

\begin{figure*}
\centering
\includegraphics[height=10cm,angle=0]{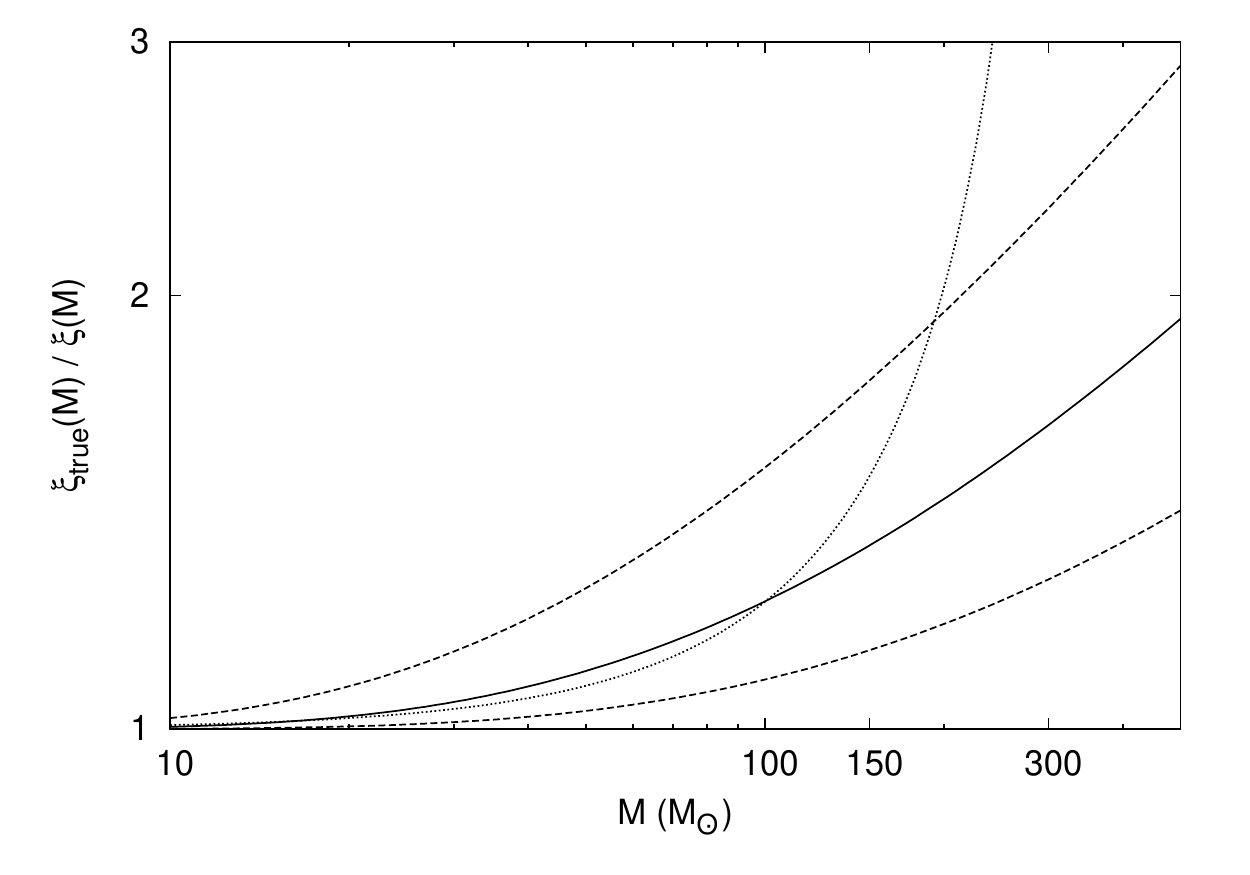}
\caption{The ratio of the ``true'' IMFs, $\xtr(M)$, in Fig.~\ref{fig:newMF} to the
``observed'' IMF, $\xi(M)$, the latter being assumed to be a
Salpeter/canonical power law  $\xi(M)\propto M^{-2.3}$.
The upper and the lower dashed, the central solid and the dotted lines correspond to
those in Fig.~\ref{fig:ejfrac1} respectively.}
\label{fig:newMFr}
\end{figure*}

It is widely believed that stars form following an initial mass function (IMF)
that has a universal form \citep{bast2010,pk2012}. In course of time however, the mass function (MF)
of a stellar population evolves as the stellar members are removed from the system.
For bound stellar clusters, the low mass stars are naturally vulnerable to removal
since they evaporate out of the system as the cluster relaxes \citep{spz,vesh97,bmk2003}. The depletion
of the low-mass stars from the bound system can be greatly augmented due to the expulsion of the residual gas
that did not form stars, in the early phase of the cluster evolution \citep{bg2007,mrk2010}. A marked
positive correlation of the global low-mass MF index ($\alpha$)
with the cluster's concentration has been
pointed out for the first time by \citet{dmarch2007}. \citet{bg2008} showed that
such a correlation can be explained by clusters born with a canonical IMF which are primordially
mass-segregated but are filling their tidal radii.
\citet{mrk2010}, however, demonstrated that the De Marchi trend can
as well be explained by residual gas expulsion from initially compact and mass-segregated clusters 
without introducing a variable low-mass IMF. This result is consistent with the compactness of freshly
formed clusters \citep{mrk2012},
whereas the tidal-radius-filling ansatz would require the clusters to condense from a cloud with
large radius matching the tidal radius.

On the other hand, the distribution of the most massive stars bound to a cluster is influenced
by the ejection of the massive members via dynamical encounters \citep{cpring92}. Massive stars in a young cluster,
which are usually in binaries, segregate in the cluster's center, where their density can be greatly
augmented. The massive stars therefore frequently interact
through super-elastic binary-single and binary-binary encounters to eject each other from the cluster,
that way depleting the massive stellar population as the cluster evolves. In general, more massive
stars are more centrally concentrated and hence are more severely depleted \citep{pf2006}. This is found to be
true in theoretical studies of massive runaway stars from star clusters by several authors
\citep{pz2011,bko2012a}.

The increasing depletion of (bound) massive stars with increasing stellar mass in star clusters has
remarkable implications on its IMF. If a young cluster is found with the massive end of its
MF in the canonical form \citep{pk2001} at its present age, then the upper end of its IMF
has to be top-heavy to achieve the canonical form at the current age (taking into account the mass loss
due to stellar evolution; see Sec.~\ref{mfevol}). In general, given a
measured MF of a young cluster near the massive limit, the corresponding stellar mass distribution
at birth, \ie, the actual IMF, has to be ``top-heavier'' depending on the age of the cluster due to
stellar-dynamical ejections.

In the present work we provide a preliminary quantification of the top-heaviness of the IMF of
a star cluster as inferred from the above considerations. To that end we utilize the ejected
fraction of massive stars, $\fej(M)$, as a function of the mass $M$ of the ejected stars from
model clusters as computed by Banerjee et al. (2012a; hereafter Paper I).
The above quantity is defined as the ratio of the number of stars, within a bin around mass $M$,
moving away from the cluster beyond $R>10$ pc from the cluster's center to the total number of stars
in the whole system, \ie, including all the bound and the ejected members, within the same mass bin.
Note that $\fej(M)$ was denoted by $g(M)$ in Paper I but we stick to this new symbol
in this paper for better clarity.
The above authors studied the runaway O-stars from the Large Megallanic Cloud (LMC) cluster
R136 by computing the dynamical evolution of realistically modelled R136-type clusters using
direct N-body integration. We point out the essentials of these computed models of Paper I
in Sec.~\ref{precomp}.

\section{Dynamical evolution of the bound stellar mass function of a star cluster}\label{mfevol}

In reality, the present-day observationally determined bound stellar mass function of a young star cluster
is corrected for unresolved multiple companions \citep{kgt91,wkm2009},
and for the wind mass loss of the stars to determine the so-called cluster IMF, $\xi(M)$.
The correction for unresolved multiple components is negligible for massive stars \citep{wkm2009}.
As pointed out in Sec.~\ref{intro}, $\xi(M)$ needs to be further corrected for the stars
removed from the bound cluster to determine the true IMF of the cluster, $\xtr(M)$,
which is the bound stellar mass function of the cluster at its birth. To correct $\xi(M)$ over the entire
mass range, one should, in principle, reassemble the cluster including all the stars that escaped
from the cluster. If the cluster is formed mass-segregated (as considered in Paper I and inferred to
be true for several young clusters; see, \eg, \citealt{bond98}), then
the massive stars reside in the cluster's center from the very beginning of its secular evolution
and they can only be ejected by dynamical encounters (\ie, they cannot be removed by the residual gas
expulsion or the external galactic field). Therefore, $\xtr(M)$ can be obtained from $\xi(M)$
by only correcting for the runaway stars, for the massive end.

In Paper I, the ejection fraction $\fej(M,t)$ is found to increase significantly with stellar mass $M$
for $M>10\Ms$ at the cluster age of $t\approx 3$ Myr (\cf Fig.~4 of Paper I). The \emph{number} of ejected stars over
the mass range $M$ to $M+dM$ is,
\begin{equation}
n_{\rm ej}(M,t)dM =\fej(M,t)\xtr(M)dM.
\label{eq:nej}
\end{equation}
Here, the decrement of the zero-age mass, $\Delta M(M,t)$, is not included since in evaluating $\fej(M,t)$,
the stellar masses at age $t$ are used for both the total population between $M$ to $M+dM$ and the
corresponding ejected population so that the same mass shift ($\Delta M$) is applicable to all the three functions
$n_{\rm ej}$, $\fej$ and $\xtr$. For brevity, we also suppress the explicit $t$ dependence in Eqn.~(\ref{eq:nej}).

The number of stars in the mass range $M$ to $M+dM$ remaining bound to the cluster is therefore,
\begin{equation}
\xtr(M)dM - n_{\rm ej}(M)dM = \xtr(M)\lbrace 1 - \fej(M)\rbrace dM,
\label{eq:nuphi}
\end{equation}
where we have used Eqn.~(\ref{eq:nej}). The right-hand-side of Eqn.~(\ref{eq:nuphi}) is
the ``observed'' (high-mass) IMF of the cluster without including the ejected stars
(but corrected for stellar evolution). In other words, 
\begin{equation}
\xi(M) = \xtr(M)\lbrace 1 - \fej(M)\rbrace,
\label{eq:nuphi2}
\end{equation}
or,
\begin{equation}
\xtr(M) = \frac{\xi(M)}{1 - \fej(M)}.
\label{eq:phirho}
\end{equation}
Eqn.~(\ref{eq:phirho}) connects the observed cluster IMF, $\xi(M)$, with its true IMF, $\xtr(M)$. 

\section{Previous computations}\label{precomp}

State-of-the-art calculations have been performed in Paper I to evolve model star clusters
that mimic R136 in terms of observable parameters. The primary objective of these calculations,
which are so far the largest sized ones with fully mass-segregated hard massive binaries (see below), was to
study the ejected OB-stars from R136, in particular, whether massive runaway stars like VFTS 682 and 30 Dor 016
can indeed be ejected dynamically from R136 as suspected \citep{blh2011,evns2010}.
These computations demonstrated good agreement of the kinematics of the massive ejecta with those of these
noted runaways. Furthermore, they suggest that the tight massive binaries that drive these runaways can merge due to
the dynamical interactions leading to the formation of ``super-canonical'' \citep{pk2012} stars,
\ie, stars with masses exceeding the canonical upper limit of $150\Ms$ (see Paper I and \citealt{bko2012b}), in agreement
with the observed stellar population in R136 \citep{crw2010}.
In this study, we continue to utilize the results of these computations.
The initial conditions and the method of these calculations are accounted in detail in Sec.~2 of Paper I which
we briefly recapitulate here.

The above computations comprise direct N-body integrations of
Plummer spheres \citep{hh2003,pk2008} with parameters conforming with those of R136. The initial mass
of the Plummer spheres is $M_{cl}(0) \approx 10^5\Ms$ which is an upper limit for R136 \citep{crw2010} and
the half mass radii are taken to be $r_h(0)\approx0.8$ pc. The clusters are made of stars with ZAMS masses
drawn from a canonical IMF \citep{pk2001} over the range $0.08\Ms < m_s < 150\Ms$ and are of metallicity
$Z=0.5Z_\odot$. With the mean stellar mass of $\approx0.6\Ms$, each cluster comprises $\approx1.7\times10^5$
stars.

As for the primordial binary population, stars with $m_s > 5\Ms$ are all in binaries while all lighter stars
are kept single. This termination of the binary population is due to computational ease; binaries bottleneck
the calculation speed of direct N-body integration significantly so that adapting a full spectrum of primordial
binaries (\ie, 100\% initial binary fraction; \citealt{mrk2012}) in models of the size that we compute becomes prohibitive. 
Disregarding the binary population for $m_s<5\Ms$ of course does not affect the ejection of massive stars and
mergers of massive binaries significantly (see Paper I for details). Following the observed period
distribution of O-star binaries \citep{sev2010}, the orbital periods of the binaries having primary masses
$m_s>20\Ms$ are chosen from a uniform distribution between $0.5<\log_{10}P<4$, where $P$ is the orbital period in
days. For primary masses $5\Ms<m_s<20\Ms$, the ``birth period distribution'' of \citet{pk95b} is adapted. 
The eccentricity distribution is chosen to be thermal \citep{spz,pk2008}. These distributions, however, are
not modified via ``eigenevolution'' \citep{pk95b} as the interactions between massive pre-main-sequence
stars are not yet quantified.
The initial configurations are fully mass-segregated \citep{bg2008} so that the massive binaries concentrate
within the clusters' central region mimicking primordial mass segregation which is inferred for
several Galactic globular clusters and young star clusters.

To perform the direct N-body computations,
the state-of-the-art ``NBODY6'' integrator \citep{ar2003} has been used. In addition to integrating the particle
orbits using the highly accurate fourth-order Hermite scheme and dealing with the diverging gravitational
forces in close encounters through regularizations, NBODY6 also employs the analytical stellar and binary
evolution recipes of \citet{hur2000,hur2002}. The models are computed for $\approx 3$ Myr four times,
each with a different random number seed initialization.

\section{Ejected fraction of massive stars $\fej(M)$: a
top-heavy IMF in R136}\label{ejfrac}

Figure 4 of Paper I shows the ejected fraction of stars as a function of stellar mass $M$, where
a significant and monotonic increase of $\fej(M)$ can be seen beyond $M> 10\Ms$. As explained in Paper I,
this trend is a direct consequence of mass segregation; more massive systems are more concentrated
towards the cluster's center and hence interact more efficiently (see below). Fig.~\ref{fig:ejfrac1} re-plots the
$\fej(M)$ at the cluster age $t\approx3$ Myr,
as obtained from the computations of Paper I, for $M>10\Ms$ where we use the same symbols as in
Figure 4 of Paper I to differentiate among the data from different computed models.

A possible form of the fitting function to the data in Fig.~\ref{fig:ejfrac1} would be asymptotic to 1 with
increasing $M$ as $\fej(M)<1$ by definition but monotonically increasing as the trend of the data suggests.   
Due to the scatter of the points in Fig.~\ref{fig:ejfrac1} over a large range and to ensure
reliability, we do a least-squares fit to the logarithm of the data in
Fig.~\ref{fig:ejfrac1}. We adopt the exponential functional form,
\begin{equation}
\log(\fej(M)) = -Be^{-px} \left(B>0,p>0\right),
\label{eq:ejidx2}
\end{equation}
to best-fit with the logarithm of the data-coordinates of Fig.~\ref{fig:ejfrac1}. This functional form
asymptotically increases to 0, \ie, approaches 1 when transformed into the linear scale.
We use all the 28 data-points
of the above figure to obtain the best fit values $B = 8.03\pm1.42$ and $p = 1.20\pm0.13$.
The central solid line in Fig.~\ref{fig:ejfrac1} represents the best-fit line to the
data and the two dashed lines represent the upper and the lower bounds of the fit corresponding to
the errors (1-$\sigma$ or the standard errors) in $B$ and $p$.

As such, the data-points in Fig.~\ref{fig:ejfrac1}
can be found to follow an overall power-law index of $\approx 1.5$ if one fits a power-law
to them. A basic physical argument that supports such an increment is as follows.
The massive single stars get ejected by super-elastically \citep{hh2003} encountering with binaries
within the cluster (as noted in Paper I, most of the ejected entities are single stars,
given the initial distribution of the binaries' orbital periods adapted here\footnote[1]{
Note that ejected single stars may in reality be binaries if the models would incorporate a large
fraction of O-stars being in very tight ($<$ few au) binaries within wider triple or quadruple systems
with outer orbital periods as chosen here (see \citealt{chin2012}).}).
These single stars are liberated from binaries either by binary-binary encounters or by direct
disruption of binaries and they can, in general, be expected to remain mass-segregated within the cluster.
A given segregated group of stars of mass $M$ remain confined within a radius $r\propto M^{-1/2}$ of the
cluster (see Chapter 16 of \citealt{hh2003}) so that their density $\propto r^{-3}\propto M^{3/2}$. This
density is also proportional to the single-stars' encounter rate with the binaries and hence to the
probability of the formers' ejection from the cluster. Hence, one can expect that the ejection fraction
$\fej(M)\propto M^{3/2}$ which is indeed close to the best-fitted power-law index as obtained here.

The dotted line in Fig.~\ref{fig:ejfrac1} represents the best-fit power law $\fej(M)=AM^k$ (truncated at
$\fej(M)=1$). By doing a
linear least-squares fit to the logarithm of the data-points, we obtain the best-fit values
$A=10^{-3.65\pm0.22}$ and $k=1.46\pm0.13$. Unfortunately, due to the sparsity of data-points
near the massive end it is presently difficult to determine the true functional behavior of $\fej(M)$
over the entire data of Fig.~\ref{fig:ejfrac1}. A larger number of such computations would help to settle
this issue.

The true IMFs, $\xtr(M)$,
corresponding to the $\fej(M)$ lines in Fig.~\ref{fig:ejfrac1},
as obtained from Eqn.~(\ref{eq:phirho}), is shown in Fig.~\ref{fig:newMF} where the observed IMF
is taken to be the canonical one, \ie, following the Salpeter form $\xi(M)\propto M^{-2.3}$ for $M>10\Ms$
(shown by the dash-dotted line).
The upper dashed, central solid, lower dashed and the dotted lines in Fig.~\ref{fig:newMF} correspond to those
in Fig.~\ref{fig:ejfrac1} respectively.
Noticeably, the IMF of R136 is indeed observed to be
very close to canonical over both low-mass (\citealt{and2009}; $1.1-10\Ms$)
and high-mass (\citealt{mhunt98}; $2.8-120\Ms$) ranges.

It can be seen in Fig.~\ref{fig:newMF} that the true IMF, $\xtr(M)$, is top-heavy
near its massive end even for $M<150\Ms$.
In other words, the computed ejection fraction of massive stars
from R136-type model clusters (in Paper I) indicates from a moderate to a substantial top-heaviness in the high-mass IMF
of R136. This point becomes more vivid in Fig.~\ref{fig:newMFr} where we plot the ratio of the true
IMF, $\xtr(M)$, to the Salpeter law $\xi(M)\propto M^{-2.3}$.

Notably, the fitted bounds to the data-points in Fig.~\ref{fig:ejfrac1} (the dashed lines) represent the effect
of statistical variation of the ejection-fraction, $\fej(M)$, between the computed cluster models. Such cluster
to cluster variation affects the top-heaviness of the inferred true IMF as can be seen
in the corresponding bounds of $\xtr(M)$ in Fig.~\ref{fig:newMF}. The inherent statistical fluctuations
can result in the inference of a marginally top-heavy to a moderately
top-heavy true IMF (\cf Fig.~\ref{fig:newMFr}).
In other words, if the present-day observed IMF in young clusters is canonical, then there has to be 
a cluster to cluster stochastic variation of the top-heaviness
of the true IMF with which they are born.
Further detailed study (\cf Sec.~\ref{discuss}) is needed to quantify this better. \citet{dib2010} has studied
the growth of massive stellar cores by gas accretion and the subsequent formation of stars
including the effects of stellar wind feedback
on the protocluster which indicates cluster-to-cluster variation of the stellar IMF (see also \citealt{dib2007}).

\section{Discussions and outlook}\label{discuss}

The above inference of top-heaviness near the upper end of the ``true'' IMF is not limited to
R136 but is true for any massive cluster that is observed to have a normal (\ie, canonical) IMF. The extent
of the top-heaviness of the true IMF for a particular cluster, of course, depends on its ejection
fraction $\fej(M)$. It is to be noted that the above analysis is only 
of the lowest order and is essentially a qualitative inference. This is because
had we started the N-body computations with the inferred top-heavy true IMF as corrected for the runaways,
$\fej(M)$ would have been larger than the presently used one which is obtained
from a canonical IMF. Therefore, in general,
the (top-heavier) true massive end of the IMF can be determined quantitatively from an
observed IMF by iteratively applying the correction in Sec.~\ref{ejfrac} for the
dynamical ejection of the massive stars which would tend to augment the top-heaviness.
Such a study would reveal relationships
between the top-heaviness of a cluster's IMF with its initial parameters such as
the total mass, the degree of primordial mass segregation and the concentration subject
to the constraint that a nearly canonical mass distribution is reached at the present day which is
true for many Galactic young clusters. Such relationships would, in turn, have
fundamental implications of massive star formation in a clustered environment \citep{bonn98a,zy2007}.

It is important to note that the top-heaviness that has been inferred above
for the massive (true) IMF is influenced by the adopted initial complete mass-segregation (see Sec.~\ref{precomp})
that provides the maximum efficiency in ejecting massive stars at young ages. However,
such an assumption of primordial mass-segregation is in accordance with the observations of
young star clusters; see Paper I and references therein (and, \eg, \citealt{bond98}).
As pointed out above, an interesting outlook would be to study the degree of top-heaviness
of the true IMF as a function of the initial degree of segregation. The timescale of segregation
of the massive stars/binaries is also shortened by the cluster's concentration. Therefore, it would
also be worthwhile to study the dependence of the effect on the cluster's initial concentration.
A physical basis of the true IMF being highly top-heavy can be found in \citet{dib2010}. In this
study of the appearance of the stellar IMF via accretion onto proto-stellar cores including wind feedback,
these authors do find a substantial flattening over the massive end of the IMF.

In Fig.~\ref{fig:ejfrac1}, the best-fit $\fej(M)$ plot extends beyond the canonical upper limit of $150\Ms$.
As explained in Paper I and \citet{bko2012b}, this is to take into account the ``super-canonical''
stars with zero-age masses between $150-300\Ms$ that form via mergers of massive binaries
due to dynamical encounters. Although the corrected IMF, $\xtr(M)$, is plotted in Fig.~\ref{fig:newMF}
for $M>150\Ms$ consistently with Fig.~\ref{fig:ejfrac1}, only $\xtr(M)$ for $M<150\Ms$ is
relevant as the cluster's IMF. 

In conclusion, our above analysis with the example of R136 indicates that taking into account the dynamical
ejection of massive stars, a young cluster with a seemingly normal or canonical IMF can
actually have formed with an IMF which is at least moderately top-heavy near its high-mass end.
This may well constitute the first ever detection of the
flattening of the IMF as predicted by \citet{dib2010}.
For completeness, we mention that by tracing the present-day constitution of globular clusters to
their initial configurations, a trend with birth density and the power-law index of the IMF has
emerged such that the IMF is increasingly top-heavy with increasing star-forming density above about 
$> 10^5 \Ms{\rm pc}^{-3}$ on a cluster-forming spatial scale \citep{mrketl2012}.
The present results thus add to the increasingly solid evidence that the IMF becomes top-heavy in star bursts.

\end{document}